\documentclass[prl,aps,superscriptaddress,twocolumn,floats,nofootinbib,showpacs]{revtex4}
\usepackage{amsmath,amssymb,graphicx}

\begin{document}

\title{Population extinction in a fluctuating environment}

\author{Alex Kamenev}
\affiliation{Department of Physics, University of Minnesota, Minneapolis,
Minnesota 55455, USA}
\author{Baruch Meerson}
\affiliation{Racah Institute of Physics, Hebrew University of
Jerusalem, Jerusalem 91904, Israel}
\author{Boris Shklovskii}
\affiliation{Department of Physics, University of Minnesota, Minneapolis,
Minnesota 55455, USA}
\affiliation{William I. Fine Theoretical Physics Institute,
University of Minnesota, Minneapolis, MN 55455}

\begin{abstract}
Environmental noise can cause an exponential reduction in the mean time to extinction (MTE) of an isolated population. We study this effect
on an example of a stochastic birth-death process with rates modulated by a colored Gaussian noise. A path integral formulation yields a transparent way of evaluating the MTE and finding the optimal realization of the environmental noise that determines the most probable path to extinction. The population-size dependence of the MTE changes from exponential in the absence of the environmental noise to a power law for a short-correlated noise and to no dependence for long-correlated noise.  We also establish the validity domains of the limits of white noise and adiabatic noise.
\end{abstract}


\maketitle

Extinction  of species after maintaining a long-lived 
self-regulating population is a dramatic instance of a large fluctuation. Its origin is in the intrinsic discreteness (``quantization") of individuals, and in the random nature of birth-death processes \cite{Gardiner,vankampen}. Understandably, the extinction phenomenon has attracted much attention from population biologists and epidemiologists \cite{population}.
As birth-death processes are intrinsically far-from-equilibrium, they are also of much interest to
physics \cite{Gardiner,vankampen}.   Birth-death processes often occur in time-varying
environments. Understanding the impact of environmental noise on the mean time to
extinction (MTE) of a species is both important \cite{assessment} and interesting. Early 
models assumed that the environmental noise, which modulates the birth and/or death rates of the species,
is delta-correlated in time \cite{Leigh,Lande}.  More
recently numerical simulations of the effects of a finite correlation time of the noise were performed by many population ecologists, see \textit{e.g.} \cite{Johst}. Not
surprisingly, the simulation results provide only a partial
understanding of the complex and rich interplay between the nonlinear
kinetics and intrinsic (demographic) stochasticity of the population on the one side and
the magnitude and spectral/correlation properties of the environmental
noise on the other.

In this Letter we formulate a  theoretical framework for this problem by considering a prototypical example of a single-species stochastic birth-death process with rates modulated by
a ``red" noise: a positively correlated Gaussian noise with given magnitude and correlation time. We evaluate the MTE analytically and find that the qualitative and
quantitative details of the \textit{exponential} reduction of the MTE by the environmental noise are very sensitive to the noise color. It was discovered by Leigh \cite{Leigh,Lande} that white environmental noise changes the dependence of the MTE on the metastable population size from an exponential to a power-law with a large exponent.  Here we show that a colored noise changes this exponent, reducing it at a fixed noise magnitude.  For a very long correlation time of the environmental noise, where we develop an adiabatic theory, the MTE becomes independent of the population size for a strong enough
noise.  We also establish the validity domains of the limits of white noise and adiabatic noise.

The distinct effect of the environmental noise on the MTE comes from special
realizations of the noise which affect the birth and/or death rate in an optimal way. The optimization involves
a statistical ``cost'' of a given variation of the reaction rates along with a ``gain'' due to a facilitated extinction.
We find that the \textit{optimal realization of noise} (ORN), which determines the \textit{least improbable path to extinction},
changes considerably as the noise correlation time is varied. For a short-correlated noise the ORN has a form of a sudden ``catastrophe'', bringing the reaction rates, for a certain period of time, to such values that cannot sustain a steady-state population. For a long-correlated noise the ORN merely gradually reduces the population size. While not directly causing extinction, it makes a fatal demographic
fluctuation much less improbable. The ORNs in different intermediate regimes (depending on the rescaled noise magnitude and correlation time) can be found numerically.

To be specific we consider a continuous-time birth-death process in a population of $n$ species with the birth rate $\lambda_n$ and death rate $\mu_n$ given by
\begin{equation}\label{rates}
    \lambda_n=\frac{n}{2}\, (\mu+r-a n)\,,\;\;\;\;\;\mu_n=\frac{n}{2}\, (\mu-r+a n)\,.
\end{equation}
For time-independent rate constants $\mu$, $r$ and $a$ (we assume $r<\mu$), this is a symmetrized version of the logistic Verhulst model: a well-studied model of population dynamics, see \textit{e.g.} \cite{Nasell}. The terms nonlinear in $n$ describe, at $a>0$, competition for resources  which limits the exponential population growth. As a result, the rate equation $\dot{\bar{n}} = r\bar{n}-a\bar{n}^2$ predicts, at $r>0$, a stable population of the average size $\bar{n}=K \equiv r/a \gg 1$ which sets in after the relaxation time $t_r=1/r$. Demographic noise, however, makes the ``stable" population {\em metastable}. The population actually goes extinct, as a large fluctuation ultimately brings it to the absorbing state $n=0$. Large fluctuations are rare and therefore statistically independent.  As a result, the long-time survival probability 
obeys Poisson's law
\begin{equation}\label{poisson}
    \sum\limits_{n=1}^\infty P_n(t) = 1-P_0(t)=e^{-t/\tau_0}\, ,
\end{equation}
where $P_n(t)$ is the probability to find $n$ individuals at time $t$, and $\tau_0$ is the MTE. It is a well-known result (that we will reproduce shortly) that $\tau_0$ scales exponentially with the average population size $K$, see \textit{e.g.} Ref. \cite{Nasell}. In the
limit of $r\ll \mu$, that we will be interested in, one obtains with exponential accuracy:
\begin{equation}\label{mte}
    \tau_0\propto \exp(rK/\mu)\,.
\end{equation}
where we have assumed $rK/\mu \gg 1$. 

\begin{figure}[ht]
\vspace{-1cm}
\includegraphics[width=0.95\columnwidth]{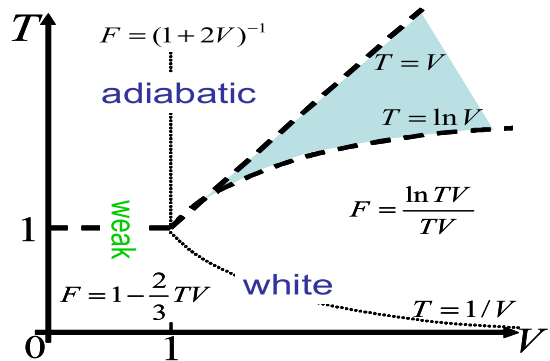}
\vspace{-.75cm}
\caption{Various regimes of extinction on the plane of rescaled parameters $V$ and $T$.
The dashed lines are schematic borders of the adiabatic, Eq.~(\ref{F-adiabatic}), and white-noise, Eq.~(\ref{F-white}) limits.
The dotted  line is the border of the weak-noise, Eq.~(\ref{F-weak}), regime. The shaded area is a crossover region.
\label{fig1}}
\end{figure}

Environmental noise manifests itself as a time-modulation of the birth and/or death rates. We will assume a modulation of the parameter $r$:
\begin{equation}\label{noise-rate}
    r\to r(t)=r - \xi(t)\,,
\end{equation}
where $\xi(t)$ is a ``red" (positively correlated)  Gaussian random process with zero mean, variance $v \ll \mu^2$ and correlation time $t_c$. For convenience, we choose the Ornstein-Uhlenbeck noise defined by the correlator $\langle \xi(t)\xi(t')\rangle =v\,e^{-|t-t'|/t_c}$. The statistical weight of a given realization of this noise is ${\cal P}[\xi(t)] \propto \exp\{-S[\xi(t)]\}$, where
\begin{equation}\label{noise}
    S[\xi(t)] = \frac{1}{4 v}\int dt\left( t_c\dot\xi^2+t_c^{-1}\xi^2\right)  \, .
\end{equation}
The environmental noise does not change the Poisson character of the survival probability, Eq.~(\ref{poisson}). Unless the noise is too weak, however, it
\textit{exponentially} reduces the MTE. We found that the MTE $\tau_\xi$, reduced by the noise, can be expressed in terms of the unperturbed MTE $\tau_0$ and two dimensionless parameters: the rescaled noise variance  $V= v K/(r \mu)$ and the rescaled noise correlation time $T=t_c/t_r=rt_c$:
\begin{equation}\label{mte-noise}
    \ln \tau_\xi= F(V, T)\, \ln \tau_0\, ,
\end{equation}
where the function $F(V, T)$ describes various  parameter regimes summarized
in Fig.~\ref{fig1}.  Importantly, each of these regimes is also characterized by an {\em optimal realization} of the noise (ORN) which causes population extinction with the highest probability. Not only the different regimes have exponentially different MTEs: they also feature qualitatively different ORNs.

Our theory, which leads to Eq.~(\ref{mte-noise}), Fig.~\ref{fig1} and other results, starts from the master equation for the time-dependent probability distribution function $P_n(t)$:
\begin{equation}\label{master}
\dot{P}_n=\lambda_{n-1} P_{n-1}- (\lambda_n+\mu_n) P_n+\mu_{n+1} P_{n+1}\,,
\end{equation}
with the birth and death rates given by Eq.~(\ref{rates}). One can show that, for $K \gg 1$ and $r \ll \mu$, this master equation can be accurately approximated by the Fokker-Planck equation, derivable by a standard procedure of van Kampen system size expansion \cite{Gardiner,vankampen}. Switching to the continuous notations $n\to q$, one can write the Fokker-Planck equation as $\dot P=\hat H P$, with the linear differential operator
\begin{equation}\label{H}
     \hat H(q,\hat p) = \frac{\mu}{2}\, \hat p^2 q + \hat p(r q -aq^2) \,.
\end{equation}
Here $\hat p =-\partial_q$ so that $[q,\hat p]=1$. In the presence of environmental noise, see Eq.~(\ref{noise-rate}), one
obtains the Hamiltonian $\hat H_\xi(q,\hat p,t) =\hat H(q,\hat p) - \xi(t)\hat p q$.

The evolution operator  $\hat U(q_f,t_f;q_i,t_i)$ of the Fokker-Planck equation can be represented
as a path integral over time-dependent trajectories $q(t)$ and $p(t)$. Below we discuss the boundary conditions
for such trajectories  for the case of population extinction.  Eventually the evolution operator must be
averaged over realizations of the environmental noise, resulting in
\begin{equation}\label{path-integral}
    \langle \hat U\rangle=\int \!{\cal D}\xi\, {\cal D}q\,{\cal D}p\,\,e^{-S[\xi]-\int \!dt\, [p\dot q-H(q,p)+ \xi pq]}\,,
\end{equation}
where $p(t)$ and $H(q,p)$ are understood as ``classical" variables, rather than the operators.

Rare events in general and population extinction in particular are described by ``classical" trajectories  accumulating a large action (and therefore having exponentially small probabilities).  For this reason the corresponding path integral can be evaluated
using the saddle point approximation near the most probable (or rather least improbable) trajectory, describing
a given rare event. Such an optimal trajectory is determined by the variation of the exponent in Eq.~(\ref{path-integral}) over
$q(t)$, $p(t)$ \textit{and} $\xi(t)$. The variation over $\xi$ yields the ORN which determines a given rare event with the highest probability. Executing this program, one arrives at the following set of classical equations of motion for $q(t)$, $p(t)$ and $\xi(t)$:
\begin{eqnarray}
  &&\dot q = \frac{\partial H}{\partial p} - \xi q\,,   \quad \quad                                             \label{q}
   \dot p = -\frac{\partial H}{\partial q} + \xi p\,,                                            \\
  && t_c^2\ddot\xi - \xi  = 2v  t_c pq \,.                                                   \label{xi}
    \end{eqnarray}
The boundary conditions, corresponding to extinction of the metastable population of average size $K$, are
$q(t=-\infty )=K\,,\, q(t=+\infty)=0$, and $\xi(t=\pm\infty)=0$. The conditions for $\xi$ follow from the fact that
the ORN must have a finite duration. Indeed, there is no need in environmental variations well before a large fluctuation starts
and well after the population goes extinct. With exponential accuracy, the extinction probability of the large fluctuation is given by
the full action, see Eq.~(\ref{path-integral}), calculated on the solution of Eqs.~(\ref{q}) and (\ref{xi}).

\begin{figure}[ht]
\vspace{-.5cm}
\includegraphics[width=0.95\columnwidth]{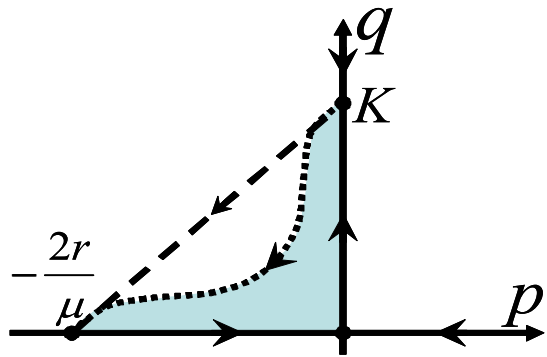}
\vspace{-.5cm}
\caption{Zero-energy trajectories of the Hamiltonian $H$ (the dashed line),  $H_v$ (the dotted line), and both $H$ and $H_v$ (the solid lines). The shadowed area is the extinction action for the short-correlated environmental noise, leading to Eq.~(\ref{F-white}).
\label{fig2}}
\end{figure}

In the absence of environmental noise, $\xi=0$,  Eqs.~(\ref{q})  admit an integral of motion: $H=const$.
Then it is easy to see that the trajectory obeying the proper boundary conditions has $H=0$ and is therefore implicitly
given by the relation $(\mu/2) p+r-aq=0$, see Eq.~(\ref{H}) and Fig.~\ref{fig2}. Calculating the action along this trajectory one finds $S=\int_K^0 p dq=r^2/(\mu a)=rK/\mu$
which yields the extinction time (\ref{mte}). Solving for $q(t)$ and $p(t)$ for this trajectory, one finds the optimal path to extinction:
\begin{equation}\label{optimal-path}
    q_0(t)=\frac{K}{e^{t/t_r}+1}\,;\quad \quad p_0(t) =\frac{-2r/\mu}{e^{-t/t_r}+1}\,.
\end{equation}
In what follows we analyze, in various  limits, the solutions of Eqs.~(\ref{q}) and (\ref{xi})
in the presence of environmental noise.

{\em Short-correlated noise.}  Here the term $t_c^2 \ddot \xi(t)$ in Eq.~(\ref{xi}) can be neglected, and the ORN becomes enslaved to the dynamics of $q$ and $p\,$: $\xi(t) \simeq -2v t_c pq$. As a result, Eqs.~(\ref{q}) become Hamiltonian equations of motion with the \textit{effective} Hamiltonian
\begin{equation}\label{H-sigma}
    H_v(q,p) =H(q,p) + v t_c p^2 q^2\, .
\end{equation}
[The same conclusion follows from the gaussian integration over ${\cal D}\xi$ in Eq.~(\ref{path-integral}) with the \textit{white-noise}
action $S[\xi]=\int\!dt \,\xi^2/(4v t_c)$.] Now, $H_v$ is an integral of motion of Eqs.~(\ref{q}) and, by virtue of the boundary conditions, it takes the value $H_v=0$.  As a result, the extinction proceeds
along the line $(\mu/2) p+r-aq+v t_c pq=0$, depicted in Fig.~\ref{fig2}. Evaluating the  action $S=\int_K^0\! pdq$ along this line, one finally arrives at Eq.~(\ref{mte-noise}) with
\begin{equation}\label{F-white}
    F(V, T)=\frac{1}{V T}
    \left[ \frac{1+ 2V T}{2V T}\, \ln(1+2V T) -1\right]\,.
\end{equation}
As the (effectively) white noise is fully characterized by the product $vt_c$, $F$ only depends on the product $VT$.

\begin{figure}[ht]
\vspace{-.5cm}
\includegraphics[width=0.95\columnwidth]{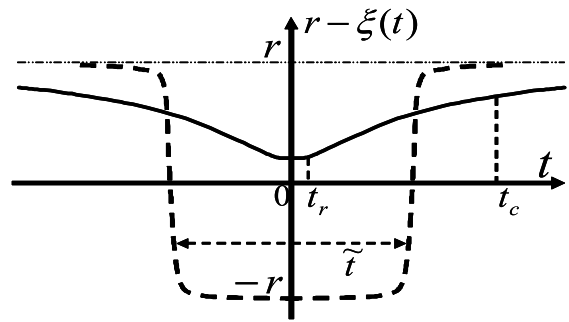}
\vspace{-.5cm}
\caption{Optimal realizations of the environmental noise in the limit of short (the dashed line) and long (the solid line) correlations of the noise. The duration of the ``catastrophe'' for the short-correlated noise is $\tilde t\approx t_r \ln (Kv t_c/\mu)$.
\label{fig3}}
\end{figure}

For a weak noise, $V T \ll 1$, Eq.~(\ref{F-white}) yields $F\simeq 1-2V T/3$. The corresponding reduction of the MTE is still exponentially large as, according to Eqs.~(\ref{mte}) and (\ref{mte-noise}),
$\tau_\xi=\tau_0e^{-2v t_c r K^2/3\mu^2}\ll \tau_0$. However, the most dramatic reduction of the MTE is predicted, in the spirit of the pure white-noise result \cite{Leigh,Lande}, in the strong-noise limit, $V T\gg 1$. Here $F\simeq \ln(V T)/(V T)$, and one obtains
\begin{equation}\label{mte-white}
    \tau_\xi\propto (v t_c K/\mu)^{r/(v t_c)} \,.
\end{equation}
One can see that the exponential scaling of the MTE with the population size $K$, \textit{cf.} Eq.~(\ref{mte}), gives way here to a power law of $K$
with a large exponent. 
To clearly see the origin of this qualitative change in the MTE, let us find the ORN leading to Eq.~(\ref{mte-white}). The logarithmic term in Eq.~(\ref{F-white}) comes from the hyperbolic part of the extinction trajectory, see Fig.~\ref{fig2}, where
$v t_c pq\simeq -r$. Here $\xi(t)\simeq-2v t_c pq \simeq 2r\simeq \mbox{const}$, and therefore $\dot q\simeq -rq$. The latter equation describes the population size decay from the initial value $K$ down to $\mu/(v t_c)$. At this scale [and at time $ \tilde t = t_r\ln (Kv t_c/\mu)$],
the demographic noise takes over the environmental one, see Eq.~(\ref{H-sigma}). As a result,  the ORN of the short-correlated environmental noise is a \textit{catastrophic} event \cite{catastrophe}, where the parameter $r>0$ suddenly
drops to $-r$ and keeps this value for a logarithmically  long time $\tilde t \gg t_r$, see Fig. \ref{fig3}. The MTE (\ref{mte-white}) merely reflects the statistical weight of this
ORN. This argument also shows that the validity of Eq.~(\ref{F-white}) requires a less restrictive condition then $t_c\ll t_r$. Indeed, it suffices to demand that $t_c\ll \tilde t =t_r \ln(V T)$, see Fig.~\ref{fig1}.

{\em Long-correlated noise.}  Here an adiabatic theory can be developed. The rare fluctuation, causing extinction, takes time about $t_r$, see Fig. \ref{fig3}.  As the environmental noise changes on a much longer time scale $t_c$, the extinction fluctuation samples an almost constant value of the noise $\xi(0)=\xi_0$, to be determined below. The \textit{effective} parameter $r$ is therefore equal to $r-\xi_0$ and is constant. Therefore, the corresponding extinction \textit{rate} is $\sim\exp[-(r-\xi_0)^2/(\mu a)]$, cf. Eq.~(\ref{mte}). Now we notice that the right hand side of Eq.~(\ref{xi}) vanishes everywhere except in a small time window $|t|\lesssim t_r\ll t_c$. As a result, the solution of Eq.~(\ref{xi}) for the ORN is $\xi(t) \simeq \xi_0e^{-|t|/t_c}$. Using it in Eq.~(\ref{noise}), we find the statistical weight of the ORN to be $\sim \exp[-\xi_0^2/(2v)]$.
Finally we need to find the optimal value of $\xi_0$ by optimizing the extinction rate against the statistical weight of the ORN. This is done by finding
the minimum of $\xi_0^2/(2v)+(r-\xi_0)^2/(\mu a)$ which is achieved at $\xi_0=r[1+\mu a/(2v)]^{-1}$.
The minimum action,
$r^2/(\mu a+2v)$, yields the logarithm of the extinction time, which is therefore given by Eq.~(\ref{mte-noise}) with
\begin{equation}\label{F-adiabatic}
    F(V,T) = (1+2V)^{-1}\,.
\end{equation}
Notice that, for a \textit{strong} long-correlated noise, $V \gg 1$ and $T \gg 1$, one obtains $\ln \tau_\xi = r^2/2v $ which is independent of the population size $K$.

When does Eq.~(\ref{F-adiabatic}) apply? It turns out that, for a strong long-correlated noise, the condition $T\gg 1$ gives way to a more restrictive one. Indeed, when deriving Eq.~(\ref{F-adiabatic}) we assumed that
$r(t)\simeq r-\xi_0 e^{-|t|/t_c}$ does not change during the relaxation time $t_r$. This requires $r^{\prime}(0)t_r\ll r(0)$ and leads to the
condition $T \gg \max(1,V)$, shown in Fig.~\ref{fig1} as the border of the adiabatic regime.

{\em Weak noise.} Here one can solve Eqs. (\ref{q}) and (\ref{xi}) perturbatively.  This is equivalent to performing the integration in Eq.~(\ref{path-integral}) over the \textit{unperturbed} extinction trajectory, Eq.~(\ref{optimal-path}) \cite{Dykman1}. The gaussian integration over the noise is done by going to the frequency space, and we obtain
\begin{equation}\label{F-weak}
    F(V, T)= 1- 4 V \int_{-\infty}^{\infty}\! \frac{d \omega}{2 \pi}\,
    \frac{(\pi  \omega)^2}{\sinh^2\pi\omega}\, \frac{T}{1+(\omega T)^2}\, .
\end{equation}
For a short-correlated noise, $T\ll 1$, this expression yields
$F=1-2V T/3$ in agreement with the limit of $V T\ll 1$ of Eq.~(\ref{F-white}). On the other hand, in the adiabatic limit, $T\gg 1$, Eq.~(\ref{F-weak}) yields $F=1-2V$ in agreement with
Eq.~(\ref{F-adiabatic}) at $V\ll 1$.   These arguments
provide the border of the weak-noise result (\ref{F-weak}), depicted in Fig.~\ref{fig1}. We stress that
even a relatively weak noise causes an \textit{exponentially} large reduction of the MTE.

Equation (\ref{F-weak}) shows that, for a weak environmental noise,  there is only one relevant scale for the noise correlation time:  $T\sim 1$. The situation is more complicated for a strong noise, $V \gg 1$. As was discussed above, the adiabatic regime holds when $T\gg V$, whereas the effectively white-noise regime holds for $T \ll \ln V$. In the crossover regime, $\ln V \lesssim T \lesssim V$ (see Fig.~\ref{fig1}), the function $F$ changes by a numerical factor of order unity. The MTE in the crossover regime can be
inferred from a numerical solution of Eqs.~(\ref{q}) and (\ref{xi}) which is quite straightforward.

To conclude, we have evaluated  the reduction of the mean time to extinction (MTE) of an isolated population caused by environmental noise. We have also established the validity domains of the limiting cases of white  and adiabatic noises. Even a relatively weak environmental noise causes an exponential reduction of MTE. A strong noise brings about qualitative changes in the scaling of  MTE with the metastable population size $K$. While  MTE scales exponentially with $K$ in the absence of environmental noise (or if the environmental noise is weak), the scaling changes to a power law in the limit of a strong short-correlated noise, and becomes $K$--independent in the limit of a strong long-correlated noise. The optimal realization of the environmental noise, which results in the population extinction with the highest likelihood, also differs qualitatively in these two limits. For a short-correlated noise the ONR has the form of a sharp ``catastrophe'' which, for a logarithmically long time, interchanges the birth and death rates of the system. For a long-correlated noise the ONR is a slow suppression of the birth rate down to a positive value. It is still debated in population biology ``whether and under
which conditions red noise increases or decreases extinction risk
compared with uncorrelated (white) noise", see Ref. \cite{Johst}(b). We hope that the analysis presented here will help resolve this and related issues.

We thank M.~Dykman for
enlightening discussions.  A.~K.  was supported by
 NSF grant DMR-0405212.  B.~M. was
supported by the Israel Science Foundation (grant No. 408/08). 


\vskip -.5cm

\begin{thebibliography}{99}

\vskip -1cm



    \bibitem{Gardiner} C.W. Gardiner, \textit{Handbook of Stochastic Methods} (Springer Verlag,
Berlin, 2004).

\bibitem{vankampen} N.G. van Kampen, \textit{Stochastic Processes in Physics and Chemistry} (North-Holland,
Amsterdam, 2001).

\bibitem{population} M.S. Bartlett, \textit{Stochastic Population Models in Ecology
and Epidemiology} (Wiley, New York, 1961).

\bibitem{assessment} S. R. Beissinger and D. R. McCullough (Editors),
\textit{Population Viability Analysis} (University of Chicago Press, Chicago, 2002).

\bibitem{Leigh} E. G. Leigh, Jr. J. Theor. Biol.  \textbf{90}, 213 (1981).

\bibitem{Lande} R. Lande, Amer. Naturalist \textbf{142}, 911 (1993).

\bibitem{Johst} (a) K. Johst and C. Wissel, Theor. Popul. Biology \textbf{52}, 91 (1997); (b) M. Schwager,  K. Johst, and F. Jeltsch, Amer. Naturalist \textbf{167} (2006).
\bibitem{Nasell} I. N{\aa}sell, J. Theor. Biol. \textbf{211}, 11 (2001).
\bibitem{catastrophe} M. Assaf, A. Kamenev and B. Meerson, arXiv:0803.0438.
\bibitem{Dykman1} A conceptually similar weak-noise theory has been recently developed by Dykman
\textit{et al.} in the context of the impact of random vaccination
on decease extinction \cite{dykmanPRL}.
\bibitem{dykmanPRL} M. Dykman, I. B. Schwartz, and A. S. Landsman, Phys. Rev. Lett. \textbf{101}, 078101 (2008).
\end{thebibliography}
\end{document}